\documentclass{article}
\usepackage{spconf,amsmath,graphicx}
\usepackage{cite}
\usepackage[english]{babel}
\usepackage{epstopdf}
\usepackage{multirow}

\title{DeviceTTS: A Small-Footprint, Fast, Stable Network for On-Device Text-to-Speech}
\name{Zhiying Huang, Hao Li, Ming Lei}
\address{Speech Lab, Alibaba Group\\
	\{zhiying.hzy, kugua.lh, lm86501\}@alibaba-inc.com}
\begin{document}
	\ninept
	\maketitle
	\begin{abstract}
		With the number of smart devices increasing, the demand for on-device text-to-speech~(TTS) increases rapidly. In recent years, many prominent End-to-End TTS methods have been proposed, and have greatly improved the quality of synthesized speech. However, to ensure the qualified speech, most TTS systems depend on large and complex neural network models, and it's hard to deploy these TTS systems on-device. In this paper, a small-footprint, fast, stable network for on-device TTS is proposed, named as DeviceTTS\footnote{Audio samples - https://zyhuang-ustc.github.io/DeviceTTS/}. DeviceTTS makes use of a duration predictor as a bridge between encoder and decoder so as to avoid the problem of words skipping and repeating in Tacotron. As we all know, model size is a key factor for on-device TTS. For DeviceTTS, Deep Feedforward Sequential Memory Network~(DFSMN) is used as the basic component. Moreover, to speed up inference, mix-resolution decoder is proposed for balance the inference speed and speech quality. Experiences are done with WORLD and LPCNet vocoder. Finally, with only 1.4 million model parameters and 0.099 GFLOPS, DeviceTTS achieves comparable performance with Tacotron and FastSpeech. As far as we know, the DeviceTTS can meet the needs of most of the devices in practical application.
		
	\end{abstract}
	
	\section{Introduction}
	\label{sec:intro}
	Text-to-Speech (TTS) aims to synthesize human-like speech from given text, and it has many applications, such as human interaction, virtual person, dubbing and entertainment. Traditional TTS can be divided into two categories: concatenative TTS and statistical parametric TTS. These two techniques are based on complex multi-stage hand-engineered pipelines~\cite{taylor2009text}, and the synthetic speech often sounds muffled and unnatural compared to the human recording. In recent years, many excellent TTS systems have been proposed to simplify the traditional TTS pipeline and remove the dependencies of hand crafted intermediate representations~\cite{wang2017tacotron, ping2018clarinet, sotelo2017char2wav, taigman2017voiceloop, li2019neural, ren2019fastspeech, yu2019durian}. After combining with neural vocoder~\cite{oord2016wavenet, kalchbrenner2018efficient, valin2019lpcnet, kumar2019melgan}, the naturalness and similarity Mean Opinion Score (MOS) are greatly improved, and even comparable with human recording~\cite{shen2018natural}. However, to make the synthesized speech more natural and high-quality, most TTS systems depend on large and complex neural network models that are difficult to train and do not allow real-time speech synthesis.
	
	With the development of artificial intelligence, more and more smart devices join our daily life. As an important part for human-computer interaction, the demand for on-device TTS increases rapidly. To build a satisfactory on-device TTS system, we first consider the feasibility of the published methods. For concatenative TTS, it depends on sufficient training dataset, the large voice package and inflexible problem make it hard to be deployed on-device. For statistical parametric TTS, it's based on two separated models and hand-set linguistic features, it's hard for model optimization to generate satisfactory synthesized speech. Tacotron\cite{wang2017tacotron}, one of the most popular End-to-End system, contains an encoder-decoder architecture and attention mechanism. The Tacotron achieves clearly better naturalness and uses a simpler training pipeline compared with traditional TTS. However, the Tacotron has the problem of words skipping and repeating when the alignment is not learned well. Furthermore, we've tried to do model compression for on-device TTS in Tacotron, but it's found that little model size leads to less robustness. This may be because that Tacotron needs enough model parameters to generate a satisfactory alignment for the attention mechanism.
	
	FastSpeech~\cite{ren2019fastspeech} is a non-autoregressive TTS network, which uses Feed-Forward Transformer blocks in the encoder and decoder. A phoneme duration predictor is introduced as the bridge between encoder and decoder so as to avoid the problem of words skipping and repeating happened in Tacotron. We try the FastSpeech using the same hyperparameters as~\cite{ren2019fastspeech}, the synthesized speech is more stable than Tacotron, but it's found that the model is short of the generalization for long utterance especially for the utterances that exceed the maximum length of the training set. We suspect that this may be due to the fact that self-attention is a global modeling method, which leads to confusion for unseen length pattern.
	
	In this paper, a small-footprint, fast, stable network for on-device TTS is proposed, and it's named DeviceTTS. Inspired by FastSpeech, DeviceTTS uses a duration predictor as the bridge between encoder and decoder so as to avoid the problem of words skipping and repeating. The basic component of DeviceTTS is Deep Feedforward Sequential Memory Network~(DFSMN)~\cite{zhang2018deep}, which is a feedforward neural network with strong modeling ability but small model size. With DFSMN blocks, DeviceTTS can generate high-quality speech with satisfactory prosody using limited model parameters. In order to speed up model prediction, multi-frame prediction is used in decoder, which generates $r~(r > 1)$ acoustic feature frames in one step. Nevertheless, the coarser grained acoustic features from multi-frame prediction may cause unnatural voice when the number of model parameters are limited. To get around this, mix-resolution decoder is proposed. In mix-resolution decoder, after getting multi-frame outputs, they are reshaped and fed to a refine network which conduct single-frame prediction. The refine network serves as a role of finer grained modeling of acoustic features. Our experiences are done with WORLD and LPCNet vocoder. Finally, with only 1.4 million~(M) model parameters and 0.099 GFLOPS, DeviceTTS achieves comparable performance with Tacotron and FastSpeech in both vocoders.
	
	\begin{figure*}[t]
		\centering
		\centerline{\includegraphics[scale=0.28]{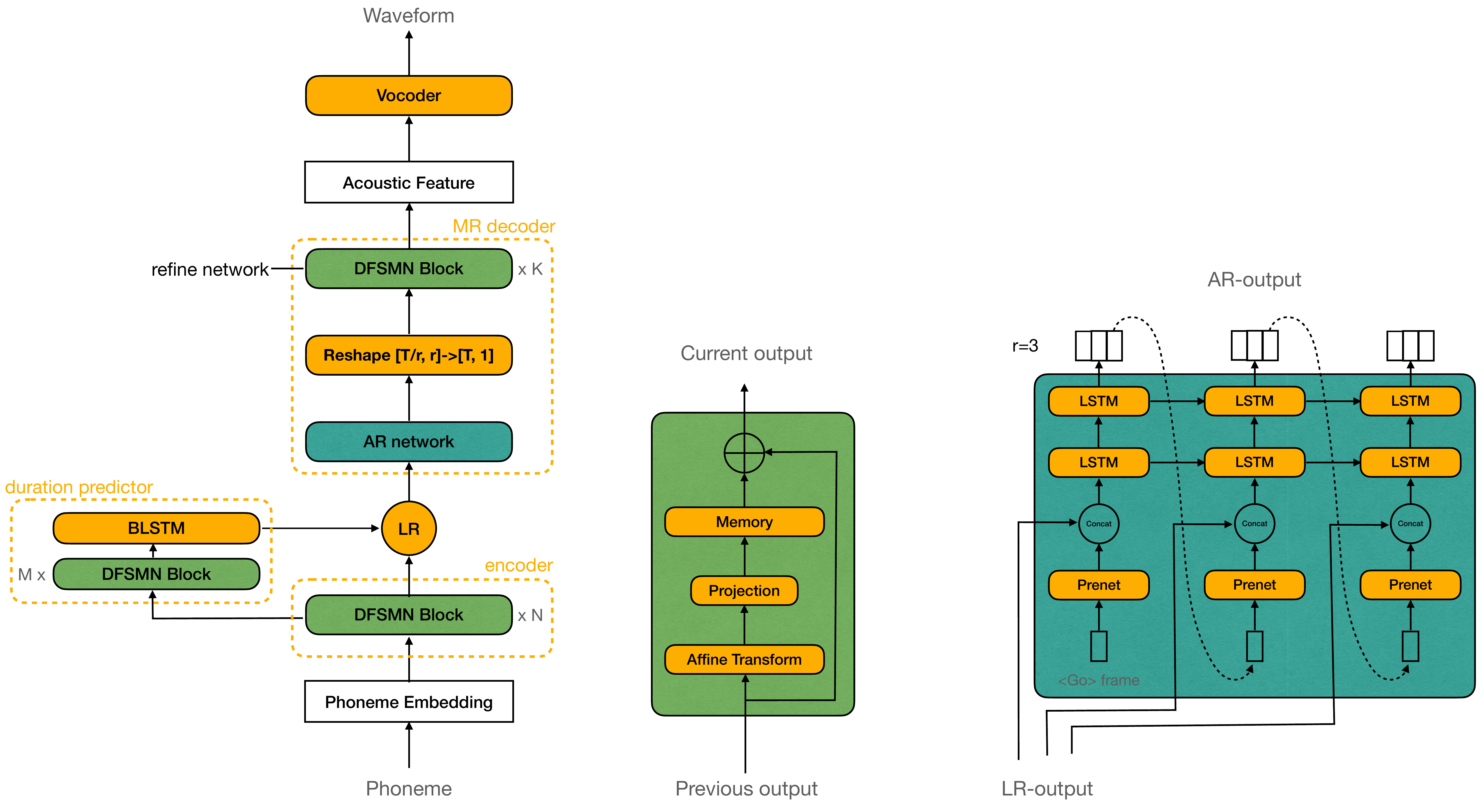}}
		\caption{The overall architecture for DeviceTTS (left), the DFSMN Block (middle), and the autoregressive~(AR) network (right). }
		\label{fig:DeviceTTS}
	\end{figure*}
	
	The remainder of this paper is organized as follows: section~\ref{sec:DeviceTTS} introduces the DeviceTTS proposed in this paper, including 1) DFSMN blocks, 2) mix-resolution decoder in detail, section~\ref{sec:Experiments} evaluates the DeviceTTS using subjective tests, and section~\ref{sec:conclusion} draws conclusions and shares our future work.

	\section{DeviceTTS}
	\label{sec:DeviceTTS}
	
	The overall architecture for DeviceTTS is shown on the left of Fig.~\ref{fig:DeviceTTS}. DeviceTTS includes four parts: encoder, duration predictor, length regulator and decoder. 
	
	The goal of the encoder is to extract robust sequential representations of text. The input is a character/phoneme sequence, where each character/phoneme is represented as a one-hot vector and embedded into a continuous vector. The encoder outputs are fed to duration predictor to get the frame number of each input character/phoneme. Then, a length regulator~(LR)~\cite{ren2019fastspeech} in the center expands encoder outputs with predicted frame number. With the expanded representations, the decoder get to generate the acoustic features.
	
	For the training of DeviceTTS, the loss function contains two parts: the acoustic features loss and the phone duration loss, as is shown in Equation (\ref{eqn:loss}). In this paper, we use Mean Absolute Error~(MAE) loss.
	\begin{eqnarray}
		\label{eqn:loss}
		\mathcal{L} = \mathcal{L}_{aco} + \mathcal{L}_{dur}
	\end{eqnarray}
	
	\subsection{DFSMN block}
	\label{subsec:DFSMN}
	As is shown, DFSMN block is the basic component for encoder, duration predictor and decoder. DFSMN is a standard feedforward neural network with memory block in the hidden layers, and the memory block is designed to encode outputs of previous hidden layer and previous histories of current layer into a fixed-sized representation. With the memory block, DFSMN can learn long-term dependency without using recurrent feedback. For training very deep neural network, skip connection between the memory blocks of successive hidden layers, where the outputs of the lower layer memory block can be directed flow to the higher layer memory block. During back-propagation, the gradients of higher layer can also be assigned directly to lower layer that help to overcome the gradient vanishing problem.
	
	The formulation of the DFSMN block takes the following form:
	\begin{eqnarray}
		p^l_t & =& f(V^{l}h^{l-1}_t + b_v^l) \\
		\tilde{h}^l_t &=& U^{l}p^l_t + b_u^{l} \\
		\label{eqn:cFSMN3}
		\hat{h}^l_t &=& \tilde{h}^l_t + \sum^{N_1}_{i=0}a^l_i\odot \tilde{h}^l_{t-i} + \sum^{N_2}_{j=1}c^l_j\odot \tilde{h}^l_{t+j} \\
		h^l_t &=& h^{l-1}_t + \hat{h}^l_t
	\end{eqnarray}
	
	$p^l_t$ denotes the outputs of Affine Transform at time $t$ while $V^l$ and $b^l$ are the corresponding weights and bias, $h^{l-1}_t$ and $h^l_t$ denote the outputs of ($l-1$)-th and $l$-th hidden layer. $\tilde{h}^l_t$ is the outputs of projection which is used for model compression. $\hat{h}^l_t$ are the outputs of current memory block with context information. In Equation (\ref{eqn:cFSMN3}), $a^l_i$ and $c^l_j$ are the look-back and look-ahead filters, and $N_1$ and $N_2$ are the orders respectively. 
	
	For DFSMN, the total latency is relevant to the look ahead filters order ($N_2$) in each memory block. When speech synthesis system is applied to some real-time applications, it is essential to control the latency. DFSMN block makes use of a latency control window size to learning the context, and the local modeling method make neural network much more stable. 
	
	\subsection{mix-resolution decoder}
	\label{subsec:AR Decoder}
	In this part, we describe the mix-resolution~(MR) decoder in detail. The MR decoder in Fig.~\ref{fig:DeviceTTS} contains two parts: AR network and refine network. AR network conduct multi-frame~(in Fig.~\ref{fig:DeviceTTS}, $r=3$) prediction autoregressively. Given LR-outputs, we select the corresponding frames by sampling at equal intervals, which is the same as the multi-frame number~$r$. Then, the selected frames are concatenated with the outputs of Prenet for the following recurrent neural network. After getting multi-frame outputs, they are reshaped and fed to a refine network which conducts single-frame prediction. The refine network serves as a role of finer grained modeling of acoustic features. In section~\ref{subsubsec:Mix-Resolution}, some experiences are done to prove the effectiveness of MR decoder.
	
	\section{Experiments}
	\label{sec:Experiments}
	
	\subsection{Speech corpus}
	\label{subsec:Speech corpus}
	
	Experiments are conducted with a corpus of female native Mandarin speaker, phonetically and prosodically rich. Speech signals are sampled at 24kHz. The speaker has approximately 30,000 utterances, with the total audio length of approximate 30 hours. This 30hs' corpus contains the prosody annotation and phone level timestamp for each utterance. So, there is no need to use align-tool to generate the phone duration as the duration predictor target, and we use the hand-set phone duration from the corpus. We think it has little effect on the overall experimental results. 
	
	\subsection{Experiment setup}
	\label{subsec:Experiment setup}
	
	\subsubsection{Model Configuration}
	
	The model hyperparameters of DeviceTTS is shown in Table~\ref{tab:Model Hyperparameters}. For DFSMN, $N$, $M$ and $K$ means the number of DFSMN blocks of encoder, duration predictor and decoder respectively, $P_1$ and $P_2$ are the hidden width of Affine Transform and Projection respectively,  $N_1$ and $N_2$ are the look-back and look-ahead orders~(see Equation (\ref{eqn:cFSMN3})).
	
	\begin{table}[t]
		\caption{Model Hyperparameters of DeviceTTS.}
		\label{tab:Model Hyperparameters}
		\centering
		\begin{tabular}{|l|l|}
			\hline
			Module & Hyperparameters \\
			\hline
			Enc/Phn\_Embd & dim=128 \\
			\hline
			Enc/DFSMN & $N$=4, $P_1$=256, $P_2$=128, $N_1$=20, $N_2$=20 \\
			\hline
			Dec/Prenet & DNN~[2x128] \\
			\hline
			Dec/RNN & LSTM~[2x128] \\
			\hline
			Dec/DFSMN &$K$=2, $P_1$=256, $P_2$=128, $N_1$=10, $N_2$=10 \\
			\hline
			Dur/DFSMN & $M$=3, $P_1$=256, $P_2$=128, $N_1$=20, $N_2$=20 \\
			\hline
			Dur/BLSTM & [1x128]~(128 is set for each direction) \\
			\hline
		\end{tabular}
	\end{table}
	
	The FastSpeech and Tacotron are both used as our baseline. For the FastSpeech, there are 6 Feed-Forward Transformer~(FFT) blocks on both the encoder and decoder side, and all the model hyperparameters are the same as the appendices of paper~\cite{ren2019fastspeech}. For the Tacotron, we use a model structure similar to the model in~\cite{shen2018natural}.
	
	\subsubsection{Vocoder}
	In this paper, we combine DeviceTTS with two kinds of vocoder, WORLD vocoder and LPCNet vocoder, to reconstruct speech waveform from predicted acoustic features. The reason for choosing these two vocoders is on the basis of high-quality speech, stability and lightweight.
	
	For WORLD vocoder, window size and window shift are 25ms and 5ms in feature extraction. We use 67-dim WORLD features, comprising 60-dim mel-cepstral coefficients, 3-dim log F0 ($static + \Delta + \Delta\Delta$), 3-dim aperiodicity of center frequencies~(CAP)~\cite{li2018emphasis} and 1-dim UV flag. Linear interpolation of F0 is done over unvoiced segments before modeling, and the WORLD features are normalized to 0-mean and unit-variance. In synthesis stage, predicted acoustic features are de-normalized and then sent to WORLD vocoder. 
	
	For LPCNet vocoder, window size and window shift are 20ms and 10ms in feature extraction. We use 23-dim LPCNet features, comprising 21-dim Bark-scale~\cite{moore2012introduction} cepstral coefficients, and 2 pitch parameters (period, correlation). 
	
	In DeviceTTS and Tacotron, they both conduct multi-frame prediction, and the hyperparameter of outputs per step $r$ should be set before training. Considering the different window shifts in WORLD and LPCNet vocoder. The $r$ is set to 8 while using WORLD vocoder, and the $r$ is set to 3 while with LPCNet vocoder.
	
	\subsubsection{Training and Test}
	We train Tacotron, FastSpeech and DeviceTTS in 1 NVIDIA V100 GPU, with batch size of 32/16/32 sentences respectively. We use the Adam optimizer with $\beta_{1}=0.9$, $\beta_{2}=0.999$, $lrate=0.002$ and $\epsilon=10^{-9}$, and follow the same learning rate schedule in~\cite{vaswani2017attention}. It takes 300k steps for training until convergence.
	
	For testing, a testing set with 100 utterances in a wide range of fields are collected for evaluation. We measure the quality of the synthetic utterances by conducting a crowd-sourced naturalness MOS test, whereby listeners in Alicrowd Platform are asked to rate the naturalness of synthetic utterances and recordings on a nine-scale score from 1 to 5 with 0.5-point increments. Each utterance is listened 5 times by different listeners, and about 200 listeners join for MOS test. As is mentioned above, FastSpeech is short of generalization for long utterance especially for the utterances that exceed the maximum length of the training set. In the following experiences, the testing utterances for FastSpeech are limited to the maximum length of the training set.
	
	\begin{table}[t]
		\caption{Comparison of naturalness Mean Opinion Score (MOS) of different systems. (95\% confidence intervals computed from the t-distribution for various systems.)}
		\label{tab:quality with WORLD Vocoder}
		\centering
		\begin{tabular}{|c|l|c|}
			\hline
			Vocoder & System & MOS \\
			\hline
			- & GT & 4.330$\pm$0.119 \\
			\hline
			\multirow{3}{*}{WORLD} & Tacotron & 4.226$\pm$0.108 \\
			\cline{2-3}
			& FastSpeech & 4.227$\pm$0.107 \\
			\cline{2-3}
			& DeviceTTS & 4.193$\pm$0.109 \\
			\hline
			\multirow{3}{*}{LPCNet} & Tacotron & 4.285$\pm$0.109 \\
			\cline{2-3}
			& FastSpeech & 4.276$\pm$0.108 \\
			\cline{2-3}
			& DeviceTTS & 4.255$\pm$0.103 \\
			\hline
		\end{tabular}
	\end{table}
	
	\begin{table}[t]
		\caption{Model Complexity~(WORLD vocoder).}
		\label{tab:Model performance}
		\centering
		\begin{tabular}{|l|c|c|}
			\hline
			System & \#Paras. (M) & GFLOPS \\
			\hline
			Tacotron & 13.5 & 0.491\\
			\hline
			FastSpeech & 36.1 & 8.58 \\
			\hline
			DeviceTTS & 1.4 & 0.099 \\
			\hline
		\end{tabular}
	\end{table}
	
	\subsection{Results}
	\label{subsec:Audio Quality}
	\subsubsection{Audio Quality}
	\label{subsubsec:Audio Quality}
	In this part, we compare the naturalness MOS of different systems with WORLD vocoder and LPCNet vocoder. The MOS result is shown in Table~\ref{tab:quality with WORLD Vocoder}.  In line with the expectations, Tacotron shows the best performance because of the soft alignment from the attention mechanism between the encoder and decoder makes the prosody of synthesized speech smoother. The FastSpeech is slightly worse than Tacotron in naturalness MOS, with average MOS gap of -0.004. Compared with Tacotron and FastSpeech, DeviceTTS shows comparable naturalness, with MOS of 4.193 for WORLD vocoder, and MOS of 4.255 for LPCNet vocoder. 

	\subsubsection{Complexity}
	\label{subsubsec:Complexity}
	Then, we compare the model complexity and computational complexity, both of them influence the usage of memory and central processing unit~(CPU) on-device. The results are illustrated in Table~\ref{tab:Model performance}, and they are based on WORLD vocoder. For model complexity, it's found that DeviceTTS has only 1.4 M model parameters while FastSpeech uses 36.1 M model parameters and 13.5 M for Tacotron. We also calculate the computational complexity of these three systems while predicting the acoustic features of an audio with 1 second, the results are shown in the third column of Table~\ref{tab:Model performance}. It's worth noting that FastSpeech has a much larger GFLOPS than Tacotron~(8.58 GFLOPS VS. 0.491 GFLOPS). The reason for that is because the $r$ is different for these two systems, the $r$ of FastSpeech is set to 1~(the same as~\cite{ren2019fastspeech}) while the $r$ is set to 8 for Tacotron. Also, it's remarkable that the GFLOPS of DeviceTTS~(0.099 GFLOPS) is much less than Tacotron~(0.491 GFLOPS) and FastSpeech~(8.58 GFLOPS).
	
	From the above results, we can see that DeviceTTS has obvious advantages in model size and inference speed comparing with Tacotron and FastSpeech. As far as we know, the DeviceTTS that with 1.4 M model parameters and 0.099 GFLOPS can meet the needs of most of the devices in practical application.
	
	\subsection{Ablation Study}
	\label{subsec:Ablation Study}
	In this section, we conduct ablation studies to verify the effectiveness of several components in DeviceTTS, including MR decoder and AR network. CMOS evaluation is used for these ablation studies, and we use WORLD vocoder to recover the waveform from predicted acoustic features.
	
	\subsubsection{MR decoder}
	\label{subsubsec:Mix-Resolution}
	The MR decoder is proposed to predict finer grained acoustic features as is described in Section~\ref{subsec:AR Decoder}. In this part, we conduct experiments to assess the impact of MR decoder. To make comparable results, DeviceTTS without MR decoder use similar number of parameters. As shown in Table~\ref{tab:CMOS MR}, removing the MR decoder results in -0.254 CMOS, which demonstrates the effectiveness of this structure.	
	
	\begin{table}[t]
		\caption{CMOS comparison in the ablation studies for MR decoder.}
		\label{tab:CMOS MR}
		\centering
		\begin{tabular}{|l|c|}
			\hline
			System & CMOS \\
			\hline
			DeviceTTS & 0 \\
			\hline
			DeviceTTS without MR decoder & -0.254 \\
			\hline
		\end{tabular}
	\end{table}
	
	\subsubsection{AR network}
	\label{subsubsec:AR Decoder}
	In the MR decoder, an AR network is used. We also try the DeviceTTS with non-AR network, where the AR network is replaced with DFSMN blocks. The naturalness MOS comparison is shown in Table~\ref{tab:CMOS AR}, and the performance of AR network and non-AR network are almost equal, with -0.003 CMOS for non-AR network. 
	
	We also calculate the GFLOPS for generating the first output frame, they are shown in the third column of Table~\ref{tab:CMOS AR}. As we can see, the DeviceTTS with AR Network has about 60\% less GFLOPS than DeviceTTS with non-AR Network. This is because that the look-ahead order~($N_2$) of non-AR network needs to be set to be large enough to keep the speech quality, so that non-AR network has larger latency than AR network.
	Therefore, AR network has obvious advantages over non-AR network.
	
	\begin{table}[t]
		\caption{CMOS comparison in the ablation studies for AR network.}
		\label{tab:CMOS AR}
		\centering
		\begin{tabular}{|l|c|c|}
			\hline
			System & CMOS & GFLOPS~(first frame) \\
			\hline
			DeviceTTS & 0 & 0.066 \\
			\hline
			DeviceTTS~(non-AR network) & -0.003 & 0.169 \\
			\hline
		\end{tabular}
	\end{table}
	
	\section{Conclusion}
	\label{sec:conclusion}
	In this paper, DeviceTTS is proposed, and it's a lightweight, fast, stable network for on-device TTS. Inspired by FastSpeech, DeviceTTS uses a duration predictor as a bridge between encoder and decoder to make the model more stable. Considering the requirement of model size and inference speed on-device, DFSMN block and MR decoder are introduced, they play an important role in DeviceTTS. Experimental results show that the DeviceTTS achieves comparable performance with Tacotron and FastSpeech with only 1.4 M model parameters and 0.099 GFLOPS. The DeviceTTS can meet the needs of most of the devices in practical application.
	
	In the future, we will try faster vocoders, such as MelGAN. Also, the performance of multi-speaker DeviceTTS will be explored.
	
	\bibliographystyle{IEEEbib}
	\bibliography{refs}
	
\end{document}